\documentclass[5p,times]{elsarticle}

\usepackage{amsmath,amssymb,amsfonts}
\usepackage{algorithmic}
\usepackage{algorithm}
\usepackage{graphicx}
\usepackage{textcomp}
\usepackage{xcolor}
\usepackage{float}
\usepackage{listings}
\usepackage{color}
\usepackage{url}
\usepackage{caption}
\usepackage{subcaption}
\usepackage{lineno,hyperref}
\usepackage{multirow}
\usepackage{tabularx}
\usepackage{parcolumns}
\usepackage{placeins} 

\definecolor{darkgreen}{RGB}{0, 153, 51}
\usepackage{subfiles} 

\begin{document}
\sloppy


\hyphenchar\font=-1
\hyphenation{persistent}
\hyphenation{metall}
\hyphenation{Metall}

\begin{frontmatter}
\title{Metall: A Persistent Memory Allocator For Data-Centric Analytics}

\author[casc]{Keita Iwabuchi\corref{ca}}
\ead{kiwabuchi@llnl.gov}

\author[casc,vt]{Karim Youssef}
\ead{karimy@vt.edu}

\author[umbc]{Kaushik Velusamy}
\ead{kaushik2@umbc.edu}

\author[casc]{Maya Gokhale}
\ead{gokhale2@llnl.gov}

\author[casc]{Roger Pearce}
\ead{rpearce@llnl.gov}

\cortext[ca]{Corresponding author}
\address[casc]{Center for Applied Scientific Computing, Lawrence Livermore National Laboratory, 7000 East Avenue, Livermore, CA 94550, United States}
\address[vt]{Department of Computer Science, Virginia Polytechnic Institute and State University, Blacksburg, VA 24061, United States}
\address[umbc]{Department of Computer Science, University of Maryland, Baltimore County, 1000 Hilltop Cir, Baltimore, MD 21250, United States}

\begin{abstract}
Data analytics applications transform raw input data into analytics-specific data structures before performing analytics.
Unfortunately, such data ingestion steps are often more expensive than analytics.
In addition, various types of NVRAM devices are already used in many HPC systems today.
Such devices will be useful for storing and reusing data structures beyond a single process life cycle.

We developed Metall, a persistent memory allocator built on top of the memory-mapped file mechanism.
Metall enables applications to transparently allocate custom C++ data structures into various types of persistent memories.
Metall incorporates a concise and high-performance memory management algorithm inspired by Supermalloc and the rich C++ interface developed by Boost.Interprocess library.

On a dynamic graph construction workload, Metall achieved up to 11.7x and 48.3x performance improvements over Boost.Interprocess and memkind (PMEM kind), respectively.
We also demonstrate Metall's high adaptability by integrating Metall into a graph processing framework, GraphBLAS Template Library.
This study's outcomes indicate that Metall will be a strong tool for accelerating future large-scale data analytics by allowing applications to leverage persistent memory efficiently.
\end{abstract}


\end{frontmatter}

\section{Introduction}

Data science has become a rapidly evolving field.
It plays an increasingly important role in science and security domains.
High volume data analytics is one of the key domains in exascale computing~\cite{ECP}~\cite{Reed2015}.
Such data analytics applications usually perform data ingestion tasks, which index and partition data with analytics-specific data structures before performing the targeted analytics.
However, the ingestion step is often more expensive than the analytics itself due to unstructured write-intensive operations on large volumes of data.
In addition, the same or derived data is re-ingested frequently in real situations -- for example, running multiple analytics to the same data with different parameters or developing/debugging a data analytics program.
An often overlooked but common theme among the variety of data analytics platforms is the need to persist data beyond a single process lifecycle.

There have been significant performance improvements and cost reductions in both software and hardware technologies of non-volatile random-access memory (NVRAM).
These devices offer cost-effective ways of persistently storing large datasets with efficient means of accessing the data for processing.

We anticipate that \emph{persistent data-centric analytics} will be a powerful model for accelerating next-generation large-scale data analytics.
In the model, applications use NVRAM as \emph{persistent memory}, i.e., applications can access data transparently using standard memory operations while the data can live beyond a single process lifecycle.

To enable the persistent data-centric analytics, we developed a persistent memory allocator, Metall\footnote{Metall is available at https://github.com/LLNL/metall}.
Metall is built on top of the memory-mapped file mechanism (mmap(2)) to allow applications to allocate and access data in persistent memory transparently.
Metall employs the rich C++ interface developed by Boost.Interprocess~\cite{Boost} so that applications can allocate custom C++ data structures in persistent memories with a small code migration cost.
Metall provides persistent memory snapshotting (versioning) capabilities.
As for the internal architecture, Metall incorporates a concise and high-performance memory management algorithm that is based on a heap memory allocator, SuperMalloc~\cite{Supermalloc}.
We also developed a user-level mmap technique, \emph{batch synchronized mmap} (\emph{bs-mmap}), to improve sparse data update performance on network-attached file systems.

The rest of this paper is structured as follows. Section~\ref{sec:preliminary} introduces preliminary knowledge of this work.
Section~\ref{sec:metall}, Section~\ref{sec:metall_internal}, and Section~\ref{sec:privateer} introduce Metall, Metall internal architecture, and bs-mmap, respectively.
Section~\ref{sec:evaluation} shows the performance of Metall and bs-mmap on dynamic graph construction workloads. Section~\ref{sec:gbtl} demonstrates Metall's high adaptability and impact by integrating Metall into a graph processing framework, GBTL~\cite{GBTL}.
Section~\ref{sec:related_work} contains related works.
Finally, Section~\ref{sec:conclusion} offers our conclusions.

In summary, our main contributions are as follows:
\begin{itemize}

\item We demonstrate the benefit of the persistent data-centric analytics model and developed a persistent memory allocator, Metall;

\item Metall is designed to allow applications to transparently allocate memory into various persistent memory devices with a reasonable code migration cost;

\item Metall exhibits up to 11.7x  and  48.3x  performance improvements over two state-of-the-art memory allocators: Boost.Interprocess~\cite{Boost} and memkind (PMEM kind)~\cite{memkind}, respectively, on a dynamic graph construction workload with node-local conventional NVMe SSD and emerging byte-addressable persistent memory (Section~\ref{sec:evaluation});

\item We present techniques to improve sparse data update performance on network-attached file systems (Section~\ref{sec:privateer});

\item We show Metall's high adaptability and impact on a real graph processing workload using a graph processing framework, GBTL~\cite{GBTL} (Section~\ref{sec:gbtl}).

\end{itemize}

\section{Preliminary}
\label{sec:preliminary}

\subsection{Persistent Memory}
There have been substantial performance improvements and cost reductions in non-volatile memory (NVRAM) technology.
For example, emerging non-volatile dual in-line memory module (NVDIMM), which is installed in the same DIMM slot as DRAM
and can provide byte-addressable accesses, is expected to play a role between DRAM and conventional NVRAM (Table~\ref{tbl:mem-perf}).
Furthermore, high-performance computing (HPC) systems have various types of NVRAM devices in production systems today, such as locally-attached devices and network-attached distributed file systems~\cite{JSFI162}.

\begin{table}
\caption{Performance comparison of memory devices}
\label{tbl:mem-perf}
\centering
\begin{tabular}{crrc}
\hline
Device & \multicolumn{1}{c}{\begin{tabular}[c]{@{}c@{}}Latency\\ (read/write)\end{tabular}} & \multicolumn{1}{c}{\begin{tabular}[c]{@{}c@{}}Bandwidth\\ (read/write)\end{tabular}} & Source \\ \hline
DDR4 DRAM & 100/100 ns & 100/37 GB/s & \cite{Hirofuchi} \\ \hline
\begin{tabular}[c]{@{}c@{}}NVDIMM\\ (Intel Optane)\end{tabular} & 370/400 ns & 38/3 GB/s & \cite{Hirofuchi} \\ \hline
PCIe NVMe SSD & 10 us & 2.5/2.2 GB/s & \cite{Lee19,OptaneSSD} \\ \hline
\end{tabular}
\end{table}

To store data into NVRAM, utilizing file systems is highly beneficial since we can support various NVRAM devices transparently and leverage existing powerful technologies to manage and move large-scale data with high reliability.
Therefore, we design Metall to work on top of a file system.

For the purposes of this paper, we use the term \emph{persistent memory} to represent a storage device/system that works with a filesystem --- including NVDIMM, NVMe SSD, and distributed file systems.

\subsection{Memory-mapped File}

Data serialization is a common technique to store data into files; however, dismantling and assembling large complex data structures is expensive in terms of performance and programming cost~\cite{Skyway}.

To avoid the cost, we leverage the memory-mapped file mechanism.
\emph{mmap(2)} is a system call that can map a file into a process's virtual memory (VM) space and provide applications with transparent access to the region --- applications can access the mapping area as if it were regular memory.

We show an example code block of mapping a file using mmap() in Code~\ref{list:mmap}.
A file is created and extended to 4096 bytes at lines 1--3.
In line 5, the file is mapped into the process's VM space with the read/write mode.
If a non-NULL address is passed to the first argument of mmap(), the kernel uses it as a hint about where to map the file.
After line 5, one can use the memory space as if it were allocated by normal memory allocation functions such as malloc().
In line 9, \emph{msync(2)} flushes dirty pages back to the filesystem and waits for the I/O to complete.
The mapping is closed at line 10 by \emph{munmap(2)}.

Actual I/O is conducted with the \emph{demand paging} mechanism --- operating systems perform I/O on-demand by page granularity and keep page cache in DRAM.
I/O could happen at any point in lines 5--10 in Code~\ref{list:mmap}.
Thanks to the demand paging, an application can also map a file bigger than the DRAM capacity.

mmap() plays an essential role in memory management and is highly useful.
However, calling it directly for each memory allocation will cause significant overhead and is not practical because A) mmap() works with at least a page granularity (e.g., 4 KB) and B) each allocation requires a new backing-file.
To provide fine-grained memories for applications, one can mitigate the overheads by building another memory allocation management layer on top of a memory-mapping region.

\begin{lstlisting}[language=C++, caption={Example of using mmap(2)}, label={list:mmap}]
int fd = open("/mnt/ssd/file", O_RDWR | O_CREAT);
int size = 4096;
ftruncate(fd, size);

char* array = (char*)mmap(NULL, size, PROT_READ | PROT_WRITE, MAP_SHARED, fd, 0);
close(fd);
array[0] = 'a';

msync(array, size, MS_SYNC);
munmap(array, size);
\end{lstlisting}

\section{Metall}
\label{sec:metall}

To take advantage of the memory-mapped file mechanism while minimizing the overheads of mmap() system call,
we propose a persistent memory allocator, called \emph{Metall}, built on top of a memory-mapped region.

In this section, we describe the key features of Metall.
We first briefly introduce Metall, followed by its API, persistence policy, snapshot capability, design choice for pointers in persistent memory, and backend data store.

\subsection{Overview of Metall}
\begin{figure}
  \centering
  \includegraphics[width=0.5\linewidth]{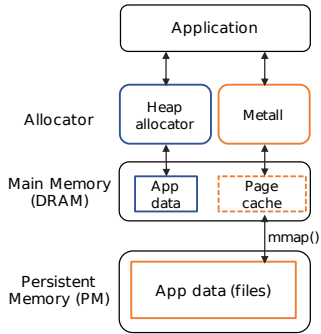}
  \caption{Data Analytics Leveraging Persistent Memory (PM)}
  \label{fig:metall_overview}
\end{figure}

Metall works on various memory devices with file system support and enables applications to allocate heap-based objects into persistent memory.
As described in Figure~\ref{fig:metall_overview}, Metall looks like a regular (heap) memory allocator from applications; however, it allocates memory into persistent memory using the memory-mapped file mechanism (mmap() system call).
Metall stores its internal memory allocation management data into persistent memory to resume memory allocation work in the subsequent execution.
Besides basic memory allocation features, Metall employs \emph{snapshot} capabilities.
Metall supports multi-threads; however, it is not designed to be shared by multiple processes, i.e., there is no interprocess synchronization support.

\subsection{Memory Allocation using Metall}
\label{sec:api}

Here we describe Metall's principal APIs, followed by two examples that allocate objects using Metall.

\subsubsection{Principal API}
Metall has C++ interfaces designed by \emph{Boost.Interprocess} (BIP)~\cite{Boost}.
Although BIP has been developed as an interprocess communication library, it has a collection of APIs useful for persistent memory allocators.
The APIs allow applications to allocate not only contiguous memory regions like malloc(3) but also complex custom data structures, including the C++ STL containers, in persistent memory.

We implemented those APIs in \emph{manager} class under \emph{metall} namespace (principal APIs are listed in Table~\ref{tbl:api}).
\mbox{allocate()} and \mbox{deallocate()} work like \mbox{malloc(3)} and \mbox{free(3)}.
\mbox{construct	\textless T 	\textgreater(char* name)} allocates \mbox{sizeof(T)} bytes and stores the address into an internal key-value store with the key ``name''.
This \mbox{construct()} function returns a proxy object whose \mbox{``() operator''} takes arguments and constructs an object of T on the allocated memory (i.e., uses the placement new) with the passed arguments.
Thus, \mbox{construct(char* name)(Args... args)} performs the multiple steps above in one line.
\mbox{find()} and \mbox{destroy()} are used to retrieve and destroy previously allocated objects by \mbox{construct()}.
\mbox{get\_allocator()} returns an object of the Standard Template Library (STL) style allocator to work with STL containers.

The manager class contains approximately fifty functions to accomplish high usability, including those with slightly different signatures\footnote{Metall API documentation: https://software.llnl.gov/metall/api/}.

\begin{table*}[t]
\caption{Metall's Principal APIs (all of them are provided by metall::manager class)}
\label{tbl:api}
\centering
\begin{tabular}{c|l}
\hline
Signature & \multicolumn{1}{c}{Description}\\
\hline \hline
void* allocate(size\_t n) &  Allocates \emph{n} bytes. \\ \hline
void deallocate(void *addr) &  Deallocates the allocated memory. \\ \hline
T* construct$\langle T, Args \rangle$(char* name)(Args... args) & Allocates and constructs an object of \emph{T} with arguments \emph{args}.\\
&Also internally stores the allocated memory address with key \emph{name}. \\ \hline
T* find$\langle T \rangle$(char* name) & Finds the already constructed object associated with key \emph{name}. \\ \hline
bool  destroy(char *name) & Destructs and deallocates the object associated with key \emph{name}. \\ \hline
metall\_stl\_allocator$\langle T \rangle$ get\_allocator$\langle T \rangle$() & Returns an STL allocator object for type \emph{T}. \\ \hline
\end{tabular}
\end{table*}

\subsubsection{Example of Memory Allocation using Metall}
An example of storing and reattaching an int object using APIs listed in Table~\ref{tbl:api} is shown in Code~\ref{list:int}.

In line 2, a Metall manager object is constructed; a backend datastore (directories and files) is created under ``/ssd/mydata'' directory, and an initial backing file is mapped to the process's virtual memory space.
In line 3, an int object is allocated and initialized with 10 (10 is passed to the constructor of the int object).
Additionally, the object's address and key (``data'') are inserted inside the key-value store in the manager object.
When the manager object is destructed (line 6), it synchronizes the allocated data with the backing files and stores its internal management data to the backing store.

In line 9, Metall opens the backing datastore created in line 2.
Metall also has a read-only open mode (metall::open\_read\_only), which protects against unintended writes
--- trying to write data will cause a segmentation fault.
In line 10, Metall searches the address of the int object allocated in line 3 from its key-value store using the key ``data''.

Finally, in line 11, the object is deconstructed and deallocated;
the corresponding entry is also removed from the internal key-value store.

\begin{lstlisting}[language=C++, caption={Example of allocating an int object using Metall}, label={list:int}]
{
 metall::manager mgr(metall::create_only, "/ssd/mydata");
 int* n = metall_mgr.construct<int>("data")(10);
 std::cout << *n; // show '10'
 *n = 20;
}
// -- Exit the program and reattach the data -- //
{
 metall::manager mgr(metall::open_only, "/ssd/mydata");
 int* n = mgr.find<int>("data").first;
 std::cout << *n; // show '20'
 mgr.destroy<int>("data");
}
\end{lstlisting}

\subsubsection{Metall with STL Container}

An STL container holds an allocator object to allocate memory storage for its elements.
Metall provides an STL-compatible allocator.
Applications can store an STL container into persistent memory by conducting the following two steps:
1) Allocate a container using Metall;
2) Pass a Metall STL allocator object to the constructor of the container object.

We show an example of storing and reattaching an STL container in Code~\ref{list:stl}.
Overall, this example is almost the same as the previous one (Code~\ref{list:int}).
In line 1, the STL-compatible allocator in Metall is passed to the vector container as the second template argument.
In line 5, an object of the vector container is allocated and constructed, receiving a Metall STL allocator object as a constructor argument.

After lines 5 and 11, as written in there, the vector object can be used transparently --- even its capacity can be changed since it holds a Metall STL allocator object internally.

\begin{lstlisting}[language=C++, caption={Example of using a STL container with Metall}, label={list:stl}]
using vector_t = vector<int, metall::allocator<int>>;
{
 metall::manager mgr(metall::create_only, "/ssd/mydata");
 auto* pvec =
 mgr.construct<vector_t>("vec")(mgr.get_allocator<int>());
 pvec->push_back(5);
}
// -- Exit the program and reattach the data -- //
{
 metall::manager mgr(metall::open_only, "/ssd/mydata");
 auto* pvec = mgr.find<vector_t>("vec").first;
 pvec->push_back(1);
}
\end{lstlisting}

\subsection{Persistence Policy}
\label{sec:persistence_policy}
Metall employs snapshot consistency, an explicit coarse-grained persistence policy in which persistence is guaranteed only when the heap is saved in a ``snapshot'' to the backing store.
The snapshot is created when the destructor or a snapshot method in Metall is invoked.
Those methods flush the application data and the internal management data in Metall to the backing store (backing files).
If an application crashes before Metall's destructor finishes successfully,
there is a possibility of inconsistency between the memory mapping and the backing files.
To protect application data from this hazard,
the application must duplicate the backing files
before reattaching the data by using either the snapshot method or a copy command in the system.

In contrast, libpmemobj in the Persistent Memory Development Kit (PMDK)~\cite{pmemio} builds on Direct Access (DAX) and is designed to provide fine-grained persistence.
Fine-grained persistence is highly useful (or almost necessary) to implement transactional object stores, leveraging new byte-addressable persistent memory fully, e.g., Intel Optane DC Persistent Memory.
However, fine-grained persistence requires fine-grained cache-line flushes to the persistent media, which
can incur an unnecessary overhead for applications that do not require such fine-grained consistency~\cite{MOD}.
It is also not possible to efficiently support such fine-grained consistency on more traditional NVMe devices.

\subsection{Snapshot}
\label{sec:snapshot}
In addition to the allocation APIs, Metall provides a snapshot feature that stores only the difference
from the previous snapshot point instead of duplicating the entire persistent heap by leveraging reflink~\cite{reflink}.

With reflink, a copied file shares the same data blocks with the existing file;
data blocks are copied only when they are modified (copy-on-write).
Because reflink is relatively new, not all filesystems support it.
The filesystems that implement reflink include \emph{XFS}, \emph{ZFS}, \emph{Btrfs}, and \emph{Apple File System (APFS)} --- we expect that more filesystems will support this feature in the future.
In case reflink is not supported by the underlying filesystem, Metall automatically falls back to a standard copy operation.

\subsection{Pointers in Persistent Memory}
\label{sec:offsetptr}
Applications have to take care of some restrictions regarding pointers to store objects in persistent memory.
Applications cannot use raw pointers for data members in data structures stored in persistent memory because there is no guarantee that backing files are mapped to the same virtual memory addresses every time.
Therefore, the \emph{offset pointer} has to be used instead of the raw pointer.
An offset pointer holds a relative offset between the address pointing at and itself so that it can always point to the same location regardless of the VM address to which it is mapped.

Metall inherits the offset\_pointer implemented in Boost.Interprocess library.
The pointer type in the STL allocator in Metall uses the offset\_pointer.
STL containers are designed to use the pointer types declared in the allocators.
Unfortunately, some implementations of containers do not work with Metall because raw pointer types are hardcoded.
Containers in Boost.Container is compatible with Metall.

Additionally, references, virtual functions, and virtual base classes have to be removed since those mechanisms also use raw pointers internally.

Other persistent memory allocators also ask applications to replace raw pointers with similar designs of offset pointers (for example, libpmemobj library in PMDK~\cite{pmemio} and Ralloc~\cite{Ralloc}).
Furthermore, the concept of the non-raw pointer is being integrated into C++ (e.g., smart pointers).
We believe that offset pointer is one of the most realistic solutions for random memory placement.
To help application developers, developing a program that assesses the compatibility of an existing data structure with Metall would be interesting future work.

\subsection{Metall Data Store}
\label{sec:datastore}
In addition to application heap data, Metall management data also have to be stored in persistent memory to resume memory allocation work.
When a Metall manager object is constructed with the create mode, it creates a root directory at the specified path;
then, the manager creates files and directories on-demand under the root directory to store its management data and application data allocated through itself.
Hereafter, we call the directory as \emph{Metall datastore} for convenience.

As all files related to a single manager are located in the same directory, one can easily duplicate or delete a Metall datastore, even using normal file copy or remove commands.

In addition, Metall is not designed for multi-process data sharing; however, multiple processes can still open the same datastore with the read-only mode.

Metall uses multiple files to store application data.
We found that breaking application data into multiple backing files increases parallel I/O performance in many situations.
When we performed a preliminary evaluation by running multi-threaded out-of-core sort,
we achieved 4.8X performance improvement by dividing the original array into 512 files (we used 96 threads and PCIe NVMe SSD).
To efficiently use persistent memory resources, Metall creates and maps new files on demand.
By default, Metall creates each file with 256 MB.
This value can be changed by defining the corresponding macro at compile time.

\section{Metall Internal Architecture}
\label{sec:metall_internal}

In this section, we explain the internal architecture of Metall.
To efficiently manage memory allocations without a complex architecture,
Metall exploits Supermalloc's main design philosophy~\cite{Supermalloc} --
virtual memory (VM) space on a 64-bit machine is relatively cheap, while physical memory remains dear.
More specifically, we take advantage of \emph{demand paging} mechanism,
that is, physical memory space both in DRAM and persistent memory is not consumed until the corresponding pages are accessed.

\subsection{Application Data Segment and Chunk}
Metall \emph{reserves} a large contiguous virtual memory (VM) space to map backing file(s) when its manager class is constructed.
Applications can set the VM reservation size when creating a new Metall datastore (Metall manager's constructor takes the value, the default size is a few TB).
Metall automatically detects the necessary VM size when opening an existing datastore.
Metall divides the reserved VM space into \emph{chunks} (2 MB by default, configurable via Metall manager's template parameter).
A chunk can hold multiple \emph{small objects} of the same allocation size (from 8B to the half chunk size).
Objects larger than the half chunk size (\emph{large objects}) use a single chunk or multiple contiguous chunks.
Metall frees DRAM and file space by chunk, that is,
small object deallocations do not free physical memory immediately, whereas large object deallocations do.

\subsection{Internal Allocation Size}
\label{sec:alloc_size}
Same as other major heap memory allocators, Metall rounds up a small object to the nearest internal allocation size.
Metall uses allocation sizes proposed by Supermalloc~\cite{Supermalloc} and jemalloc~\cite{jemalloc}.
Thanks to this approach, Metall can keep internal fragmentations equal to or less than 25 \% and convert a small object size to the corresponding internal allocation size quickly.
Metall also assigns a \emph{bin number} for each internal allocation size.
The allocation size techniques enable Metall to compute a bin number from an internal allocation size efficiently.

On the other hand, a large object (larger than 1 MB by default) is rounded up to the nearest power of 2
--- although this strategy wastes VM space, it does not waste physical memory thanks to the demand paging mechanism.
In the worst case, 1.6\% of physical memory is wasted when (1 M + 1) bytes of allocation is requested on a 4 KB page size system.
On a 64 KB page size system, 6.3\% is wasted for a (1 M + 1) bytes of allocation.

\subsection{Management Data}
\label{sec:management_data}

Metall uses three types of management data directories to manage memory allocation.
Because updating the management data causes fine-grained random memory accesses,
Metall constructs them in DRAM to increase data locality --- consequently, Metall rarely touches persistent memory when allocating memory.
Metall deserializes/serializes the management data from/to files when its constructor/destructor is called.
The cost of the process is often negligible since the management data is much smaller than the application data.

The three data directories are allocated for each Metall manager object so that multiple Metall manager objects can coexist with one another in the same program.
Here, we describe the details of the three management data directories.
\begin{figure}[t]
  \includegraphics[width=\linewidth]{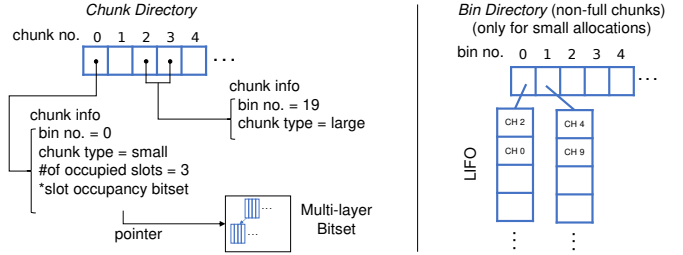}
  \caption{Metall Memory Allocation Management Data}
  \label{fig:metall_internal}
\end{figure}

\subsubsection{Chunk Directory}
The chunk directory is an array of blocks (the left figure in Figure~\ref{fig:metall_internal}).
The $i$-th block holds the status of the $i$-th chunk of the application data segment, such as bin number (internal allocation size id), chunk type (represents small or large allocation), and a pointer to a bit set for small allocation. The size of a single block is 14 bytes.

Metall utilizes a compact multi-layer bitset table and built-in bit operation functions to manage available slots in a chunk used for a small size. It can manage up to $64^{3} (=2^{18})$ slots using a three-layer structure, which is equal to the maximum number of slots if the minimum allocation size is 8B and the chunk size is 2 MB ($2^{21} / 2^3$ = $2^{18}$). Therefore, Metall calls a built-in bit operation function at most three times to find an available slot in a chunk.

Metall sequentially probes the array when it needs to find empty chunk(s).
Although we have not seen a performance bottleneck during the step,
it will be straightforward to implement an additional index structure to increase the performance.

\subsubsection{Bin Directory}
The bin directory is an array of \emph{bins} (the right figure in Figure~\ref{fig:metall_internal}).
A bin holds IDs of non-full chunks of the same internal allocation size.
A bin operates in a LIFO (last in, first out) manner.
Metall first checks this directory to find available chunks for small allocations.
If a bin is empty, Metall accesses the chunk directory to find an empty chunk.
Metall uses this directory only for small allocations since large allocations do not share chunks.

\subsubsection{Name Directory}
The name directory is a simple key-value table.
When an object is constructed by construct() function in Metall manager (Table\ref{tbl:api}),
some attributes (e.g., key string and address) of the object are stored here.


\subsection{Management Data and STL Allocator}

An STL container object holds an allocator object internally so that it can change its capacity dynamically.
To perform memory allocation work, an object of the STL allocator in Metall needs to know the address of the corresponding management data allocated in DRAM.
However, Metall cannot embed the address of the corresponding management data into an STL allocator object since the management data will not be allocated to the same VM addresses always.
In order to address this issue, a Metall STL allocator object holds the offset to the head of the application data segment using the offset pointer.
When a Metall manager object is constructed, it writes the address of its management data at the head of the corresponding application data segment.
Consequently, Metall STL containers can access the management data always.

\subsection{Multi-thread Support}
\label{sec:multi-thread}

Metall works with multiple threads.
Here, we describe how the multi-thread support is implemented in Metall.

\subsubsection{Mutex in Global Management Data}
Metall allocates a single mutex object for the chunk directory and the name directory each.

Metall also arranges a mutex object per bin in the bin directory.
As Metall does not mix different allocation sizes within a chunk, it can handle different small size allocation/deallocation requests concurrently except the following two situations:
\begin{itemize}
\item There is no entry (non-full chunk ID of the allocation size) in the bin directory during an allocation operation; thus, Metall needs to find an empty chunk in the chunk directory.
\item The last slot of a chunk has been freed. Metall needs to update the metadata of the chunk directory.
\end{itemize}

\subsubsection{Local Object Cache}
To increase multi-thread performance, memory allocators often employ local object caches.
An object cache holds recently deallocated objects.
Object caches are allocated at, for example, thread level, CPU core level, and/or CPU socket level.

Since Metall is designed to deal with larger data than existing memory allocators,
we decided to employ free-object caches at the CPU core level only to simplify its implementation.


\section{Batch Synchronized mmap (bs-mmap)}
\label{sec:privateer}

When an application maps a file using mmap(2), \emph{shared mapping} (MAP\_SHARED option) is usually used.
The file-backed shared mapping writes back updates (dirty pages) with page granularity into the underlying file system on demand.
This feature is necessary for mapping data larger than the DRAM capacity of the system (out-of-core processing).
On the other hand, network file systems such as Lustre~\cite{Lustre} are not designed to handle small and random I/Os with low concurrency~\cite{uselton2013file}.
Therefore, if applications do not need out-of-core processing with network file systems, such on-demand I/O patterns will cause unnecessary performance degradation.

A naive solution for the problem would be \mbox{\emph{data staging}}.
Specifically, 1) an application copies all files into a local memory device; 2) maps the files and performs analytics; 3) copies back to the original storage after the analytics.
However, this data staging approach could be wasteful if applications want to update data sparsely.

Another technique to mitigate the performance degradation is tuning up the behavior of the page cache by writing values to some files in /proc/sys/vm.
However, it causes 1) system-wide changes; 2) requires privilege access, which is unavailable in many large-scale clusters.

Considering these options, we designed \emph{bs-mmap}, batch synchronized mmap.
bs-mmap is a user-space file-backed memory mapping mechanism that efficiently writes back dirty pages to the backing file only when it is invoked by the application explicitly.

\subsection{bs-mmap Implementation}

bs-mmap calls mmap() with the MAP\_PRIVATE option.
MAP\_PRIVATE creates a private copy-on-write mapping where updates are not written back to the backing file by the operating system.

\emph{msync(2)} is used with the shared mapping to flush dirty pages into the backing file explicitly.
However, msync() does not work with private mapping.
Therefore, we implemented a user-level msync() that works with the private mapping.
To detect dirty pages, we used the information provided by the /proc file system on Linux systems.
The /proc file system provides an interface called \emph{pagemap}, which contains page table information about every page in a process's virtual memory space.
This information is stored in a file named \mbox{/proc/self/pagemap}, which contains a 64-bit value for each page that belongs to the process \cite{LinuxKernelPagemap}.
In the case of a private mapping, a page is no longer file-backed once it becomes dirty; however, its status is either \emph{present} or  \emph{swapped}.
Hence, a dirty page of a MAP\_PRIVATE region can be identified by checking if bit number 61 of its pagemap entry is zero and the logical OR of bits 62 and 63 equals one.
By querying these values, our msync (write-back) method can identify dirty pages without making any change in system calls or kernels.

\subsection{Bandwidth and Parallelism Utilization}
We implemented two optimizations to efficiently utilize the bandwidth and parallelism on parallel file systems.
First, bs-mmap writes back dirty pages in consecutive chunks when possible rather than page-by-page.
Second, bs-mmap writes back dirty pages in parallel. As described in Section~\ref{sec:datastore}, Metall uses multiple backing files for the application data segment. When bs-mmap flushes dirty pages using its msync() function, it assigns a thread per file to perform parallel I/O.

\section{Evaluation}
\label{sec:evaluation}

To evaluate the memory allocation performance, we perform a multi-threaded dynamic graph construction benchmark.
We also demonstrate the impact of the batch synchronized mmap technique (bs-mmap) described in Section~\ref{sec:privateer}.
The benchmark inserts edges into a graph data structure allocated in persistent memory.

\subsection{Graph Data Structure}
\label{sec:adj-list}
\begin{figure}
  \centering
  \includegraphics[width=\linewidth]{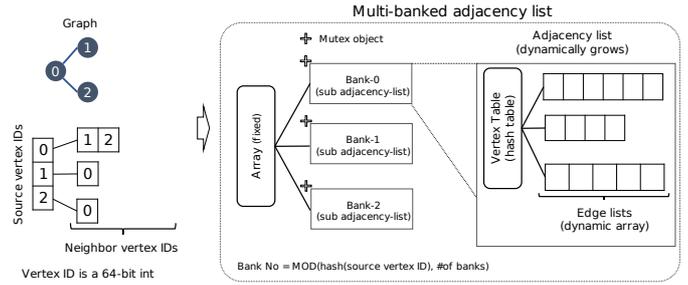}
  \caption{Banked Adjacency List Data Structure}
  \label{fig:adjlist}
\end{figure}

To construct graph data with multiple threads on a shared-memory system, we used a multi-bank \emph{adjacency list} (Figure~\ref{fig:adjlist}).
The adjacency list is one of the de facto standard graph data structures and consists of a vertex table and an edge list per vertex in the graph.
Each element in the vertex table contains the ID of a vertex and an edge list.
An edge list contains the IDs of all neighbor vertices of a vertex.
To quickly locate a specific vertex by its ID, we used the unordered\_map (hash table) container for the vertex table.
As for the edge list, we used the vector (dynamic array) container.
We used 64 bits to represent a vertex ID.

To support multi-thread graph construction, we used $m$ \emph{banks}, where $m$ is greater than the number of threads ($m = 1024$ in this experiment).
A bank is a pair of an adjacency list and a mutex object.
We constructed a graph by repeatedly inserting edges.
Each edge is a pair of source and neighbor vertex IDs, and we acquired the mutex of the bank associated with the source vertex when ingesting an edge.

To make a data structure that works with a custom STL allocator, we followed the C++ standard procedure of developing an allocator-aware class.
Precisely, we customized the multi-bank adjacency list data structure to take an allocator type in its template and an allocator object in its constructor.

\subsection{Machine Configuration}
We used three single node machines at Lawrence Livermore National Laboratory.
We show the specification of the machines in Table~\ref{tbl:machines}.

\paragraph{EPYC}
EPYC has a PCIe NVMe SSD.
We tuned the behavior of the page cache by writing values to some files in /proc/sys/vm to reduce the number of forced write-backs to the SSD device.
Specifically, we set dirty\_ratio to 90, dirty\_background\_ratio to 80, and dirty\_expire\_centisecs to a large number so that dirty pages are not evicted due to long stays in the cache.
When we performed a preliminary evaluation, we achieved significant performance improvement (up to 7X) on the graph construction benchmark.

\paragraph{Optane}
A single Intel Optane DC Persistent Memory device is installed in its DIMM slots (one side of a NUMA node) and configured with App Direct Mode.
In the App Direct Mode, the device shows up in the system as if it were a conventional block device;
we set up the device with ext4 filesystem DAX mode to bypass the page cache layer and to enable fine-grained I/O rather than page granularity.

\paragraph{Corona}
We also use one of the nodes of the Corona cluster, which consists of over 200 compute nodes.
Corona is connected to two parallel file systems: Lustre~\cite{Lustre} and VAST~\cite{VAST}.
In our environment, Lustre is suitable for large-chunk I/O and possesses higher bandwidth over VAST.
On the other hand,  VAST shows better performance for fine-grained I/O.

\begin{table}[h]
\caption{Machine Configurations}
\centering

\begin{subtable}{0.9\linewidth}
    \caption{EPYC: NVMe SSD}
    \begin{center}
        \begin{tabular}{cc}
        \hline
        CPU & \begin{tabular}[c]{@{}c@{}}AMD EPYC 7401\\ 48 cores, 96 threads\end{tabular} \\ \hline
        DRAM & 256 GB \\ \hline
        Storage & \begin{tabular}[c]{@{}c@{}}PCIe NVMe SSD 3 TB\\ Ultrastar SN200 HH-HL add-in card\\ XFS filesystem\end{tabular} \\ \hline
        Kernel & Linux Kernel v5.6 \\ \hline
        \end{tabular}
    \end{center}
    \label{tbl:epyc}
\end{subtable}

\centering
\begin{subtable}{0.9\linewidth}
    \caption{Optane: NVDIMM}
    \begin{center}
        \begin{tabular}{cc}
        \hline
        CPU & \begin{tabular}[c]{@{}c@{}}Intel Xeon Platinum 8260L\\ 48 cores, 96 threads\end{tabular} \\ \hline
        DRAM & 192 GB \\ \hline
        Storage & \begin{tabular}[c]{@{}c@{}}Intel Optane DC Persistent Memory\\1.5 TB, App Direct Mode, ext4 DAX\end{tabular} \\ \hline
        Kernel & Linux Kernel v5.9 \\ \hline
        \end{tabular}
    \end{center}
    \label{tbl:optane}
\end{subtable}

\centering
\begin{subtable}{0.9\linewidth}
    \caption{Corona: NVMe SSD and parallel file system (PFS)}
    \begin{center}
        \begin{tabular}{cc}
        \hline
        CPU & \begin{tabular}[c]{@{}c@{}}AMD EPYC 7401\\ 48 cores, 96 threads\end{tabular} \\ \hline
        DRAM & 251 GB \\ \hline
        Local Storage & \begin{tabular}[c]{@{}c@{}}PCIe NVMe SSD 1.6 TB\\ XFS filesystem\end{tabular} \\ \hline
        PFS & \begin{tabular}[c]{@{}c@{}}Lustre\end{tabular} \\ \hline
        PFS & \begin{tabular}[c]{@{}c@{}}VAST, connecting via Ethernet\\4 $\times$ 20 Gbps links\end{tabular} \\ \hline
        Kernel & Linux Kernel v3.10 \\ \hline
        \end{tabular}
    \end{center}
    \label{tbl:corona}
\end{subtable}

\label{tbl:machines}
\end{table}

\subsection{Dynamic Graph Construction}
We ran the dynamic graph construction benchmark on EPYC and Optane machines, which have node-local persistent memory, varying the sizes of input data.

\subsubsection{Implementations For Performance Comparison}
For performance comparison, we used three allocators as below:

\paragraph{Boost.Interprocess (BIP)}
Even though Boost.Interprocess~\cite{Boost} has been developed as an interprocess communication library; its managed mapped file version can work as a persistent memory allocator.
This allocator does not free space in files.
We used Boost libraries v1.75.0.

\paragraph{PMEM kind}
memkind library~\cite{memkind} provides a file-backed memory allocator (called \emph{PMEM kind}) built on top of a state-of-the-art heap allocator, jemalloc~\cite{jemalloc}.
Although PMEM kind allocates memory into a file,
it uses persistent memory as volatile memory --- i.e., it cannot reattach data or resume memory allocation beyond a single process lifecycle.
We used memkind v1.11.0.

On the Optane machine, we made a small change to this allocator because we noticed vital performance degradation due to frequently calling madvise(2) system call with MADV\_REMOVE flag to free space in both DRAM and file system.
Thus, we switched to use MADV\_DONTNEED flag to free only pages in DRAM.
Metall also uses the system call with MADV\_REMOVE for the same purpose; however, we did not make any change to Metall because Metall is designed to call the system call less frequently.

\emph{libvmem} is a similar memory allocator library included in Persistent Memory Development Kit (PMDK)~\cite{pmemio};
however, we used PMEM kind as PMDK recommends it over libvmem.

\paragraph{Ralloc}
Ralloc~\cite{Ralloc} is a persistent lock-free allocator designed and optimized for byte-addressable NVRAM (e.g., Intel Optane DC Persistent Memory).
Ralloc showed notably better performance over libpmemobj in PMDK~\cite{Ralloc}.
As Ralloc is targeted for byte-addressable NVRAM, we used it only on the Optane machine.
We used the version that was available at the time of writing.
For the purpose of this evaluation, we wrote an STL-compatible allocator class that uses Ralloc internally.

\subsubsection{Dataset}
We used an R-MAT~\cite{R-MAT} generator with the settings used in the Graph500 to generate synthetic scale-free graphs of different sizes.
We generated SCALE 24--30 graphs, scrambling the vertex IDs in order to remove unexpected localities.
The number of vertices and undirected edges in a SCALE $s$ graph are $2^s$ and $2^s \times 16$, respectively.
We treat generated edges as undirected ones; hence, the number of actually inserted edges is $(2^s) \times 16 \times 2$.
At each iteration of the benchmark, the benchmark program generates a chunk of edges into DRAM first and inserts the edges into a graph data structure. We exclude the edge generation time from reports.

\subsubsection{Results}
\begin{figure}
    \centering
    \begin{subfigure}[b]{0.9\linewidth}
        \centering
        \includegraphics[width=\textwidth]{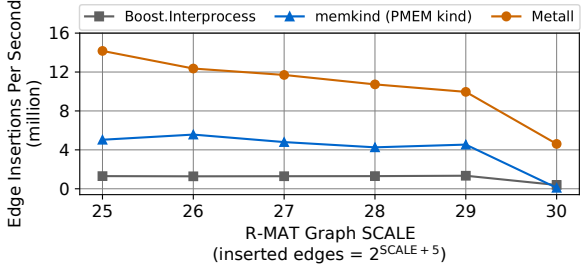}
        \caption{EPYC Machine (NVMe SSD)}
        \label{fig:const_nvme}
    \end{subfigure}
    \hfill
    \begin{subfigure}[b]{0.9\linewidth}
        \centering
        \includegraphics[width=\textwidth]{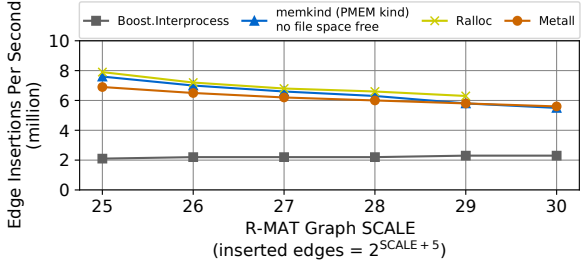}
        \caption{Optane Machine (Intel Optane DC Persistent Memory)}
        \label{fig:const_optane}
    \end{subfigure}
    \caption{The results of the multi-threaded shared-memory dynamic graph construction benchmark on two persistent memory devices. We did not run Ralloc on EPYC machine because it is designed for byte-addressable memory.}
    \label{fig:construction}
\end{figure}

We show results on the EPYC machine and the Optane machine in Figure~\ref{fig:const_optane} and Figure~\ref{fig:const_nvme}, respectively.

\paragraph{On EPYC machine}
Metall showed up to 7.4--10.9x and 2.2--2.8x improvements over Boost.Interprocess (BIP) and PMEM kind, respectively, at SCALE 25--29.
At SCALE 30, we observed performance drops in all implementations because the adjacently-list objects exceeded the DRAM capacity.
At the SCALE, Metall achieved 11.7x and 48.3x better performance over BIP and PMEM kind, respectively.

\paragraph{On Optane machine}
Metall achieved 2.1--2.3x better performance over BIP.
Ralloc did not finish at SCALE 30 because it ran out of the persistent memory space.
Metall showed similar performance to Ralloc and the modified version of PMEM kind; specifically, PMEM kind and Ralloc were up to 10\% and 14\% better than Metall, respectively.

\paragraph{Summary}
We attribute the low performance of Boost.Interprocess to its internal architecture
--- it employs a single tree with a single lock for governing memory allocation, which will not scale well with multiple threads.
Although Metall employs a simpler internal design than PMEM kind (which is based on jemalloc), it was able to achieve comparable memory allocation performance on Optane machines thanks to the design strategies proposed by Supermalloc (Section~\ref{sec:metall_internal}), such as leveraging the demand page mechanism.
Metall showed 1) the best performance on EPYC (conventional NVRAM) and 2) the compatible performance against PMEM kind and Ralloc on Optane (emerging byte-addressable NVRAM). This evaluation demonstrated Metall's high portability.

\subsection{bs-mmap on Network File Systems}
We evaluated the performance of Metall with our bs-mmap (batch synchronized mmap) described in Section~\ref{sec:privateer}, using an incremental graph construction benchmark.

\subsubsection{Benchmark Workload: Incremental Graph Construction}
The incremental benchmark uses Metall to construct a persistent graph data incrementally --- the actual data structure is the multi-bank adjacency list described in Section~\ref{sec:adj-list}.
We ran our experiments on one compute node of the Corona cluster (table~\ref{tbl:corona}).
We evaluated the performance of constructing graphs on two different network file systems, Lustre and VAST.

We used real temporal graph datasets extracted from Wikipedia and Reddit (see details in Section~\ref{sec:wiki-reddit-dataset}).
We sorted the edges of the Wikipedia and the Reddit datasets by timestamp to simulate a real-world incremental graph growing.
We partitioned each of the sorted datasets by month to iteratively and incrementally constructed a persistent banked adjacency list using Metall. The incremental benchmark's first iteration creates a new Metall datastore, adds the first chunk of edges, flushes data back to the backing store, then closes the Metall data store.
Each subsequent iteration opens the existing datastore, appends the next chunk of edges, flushes, then closes the Metall datastore.
We measured the total time per iteration, i.e., adding a monthly chunk of edges.
We broke down the measured time into ingestion time and flush time.

\subsubsection{Datasets}
\label{sec:wiki-reddit-dataset}
We used two real-world datasets in our experiments: Wikipedia page reference graph and Reddit author-author graph:

\paragraph{Wikipedia page reference graph}
We curated the Wikipedia dataset by extracting hyperlinks between all pages in the English Wikipedia dump~\footnote{https://dumps.wikimedia.org/enwiki/} as of July 1st, 2017.
The dump data contains the entire edit history of the pages in English Wikipedia from January 15th, 2001, which is the date English Wikipedia was founded.
The graph contains hyperlinks not only between article pages but also other types of pages such as author (user) pages and Category pages.
Specifically, it contains 1.8 billion hyperlink (edge) insertions.

\paragraph{Reddit author-author graph}
Reddit~\footnote{https://www.reddit.com/} is one of the largest social news websites in the world. On Reddit, users can comment on other comments. We extracted the user activities to construct an author-author comment graph.
For example, if \emph{Alice} posts a comment to \emph{Bob}'s comment, we represent it as an edge from \emph{Alice} to \emph{Bob}.
This dataset contains 4.4 billion comment activities (edges).

\subsubsection{Implementations}
We compared the performance of our \emph{bs-mmap} to that of two models that use the standard file-backed mmap (shared mapping with system \emph{msync}) as follows:

\paragraph{direct-mmap}
The first configuration consisted of mapping files directly from Lustre or VAST into Metall's virtual memory space. We considered this method as our baseline for performance comparison.

\paragraph{staging-mmap}
The second configuration brings a Metall datastore into \emph{tmpfs} and maps files from there into the virtual memory of the process in order to increase data locality during the graph construction.
tmpfs is temporary file storage configured on top of DRAM.
The staging step copies a datastore from Lustre or VAST into tmpfs at the beginning of each iteration, then copies it back at the end of each iteration.
We implemented parallel file copy-in and copy-out operations to maximize resources utilization.

\paragraph{bs-mmap}
We configured bs-mmap to read the mapped file ahead into virtual memory using~\emph{mmap}'s MAP\_POPULATE flag since this showed to be significantly faster than on-demand paging on both Lustre and VAST.
Finally, we disabled the feature of freeing file space in Metall since our preliminary experiments showed that it was an expensive operation on Lustre.
While it did not cause significant performance degradation on VAST, we disabled it in order to compare the performance across both file systems under similar conditions.

\subsubsection{Results}
\begin{figure}
    \centering
    \begin{subfigure}{0.90\linewidth}
        \centering
        \includegraphics[width=\textwidth]{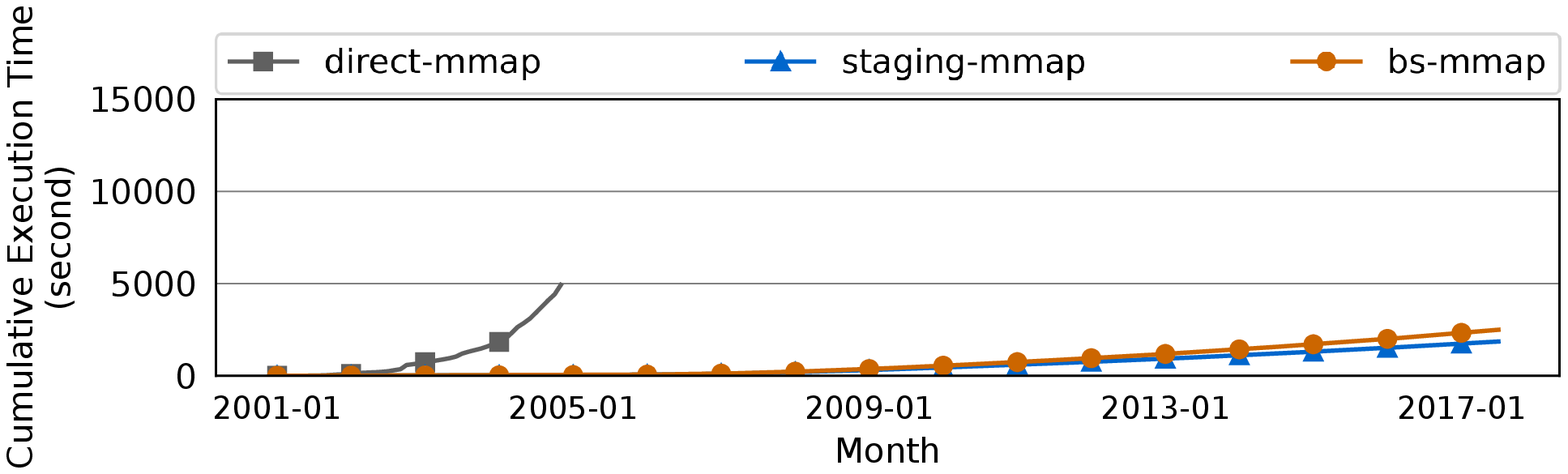}
        \caption{Wikipedia on Lustre}
        \label{fig:wikipedia_lustre}
    \end{subfigure}

    \begin{subfigure}{0.90\linewidth}
        \centering
        \includegraphics[width=\textwidth]{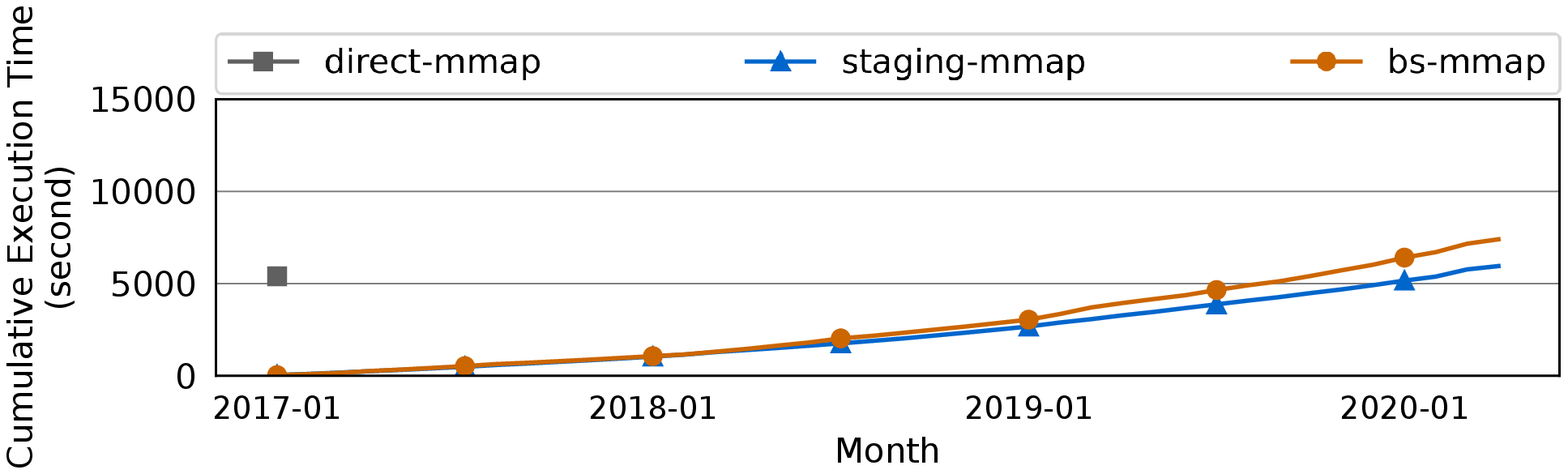}
        \caption{Reddit on Lustre}
        \label{fig:reddit_lustre}
    \end{subfigure}

    \begin{subfigure}{0.90\linewidth}
        \centering
        \includegraphics[width=\textwidth]{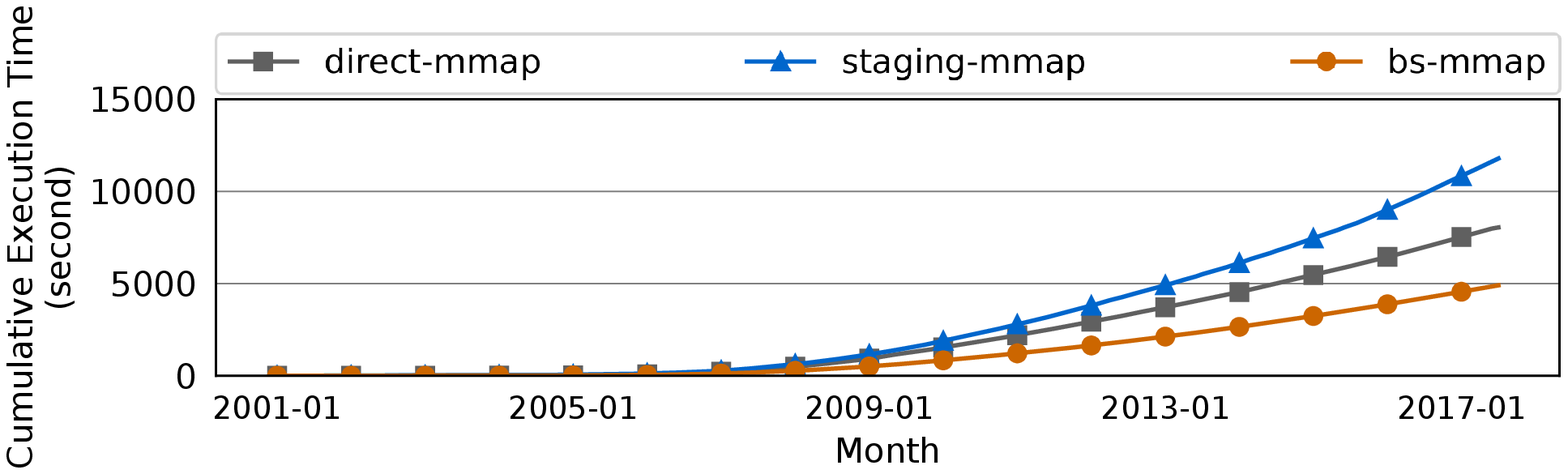}
        \caption{Wikipedia on VAST}
        \label{fig:wikipedia_vast}
    \end{subfigure}

    \begin{subfigure}{0.90\linewidth}
        \centering
        \includegraphics[width=\textwidth]{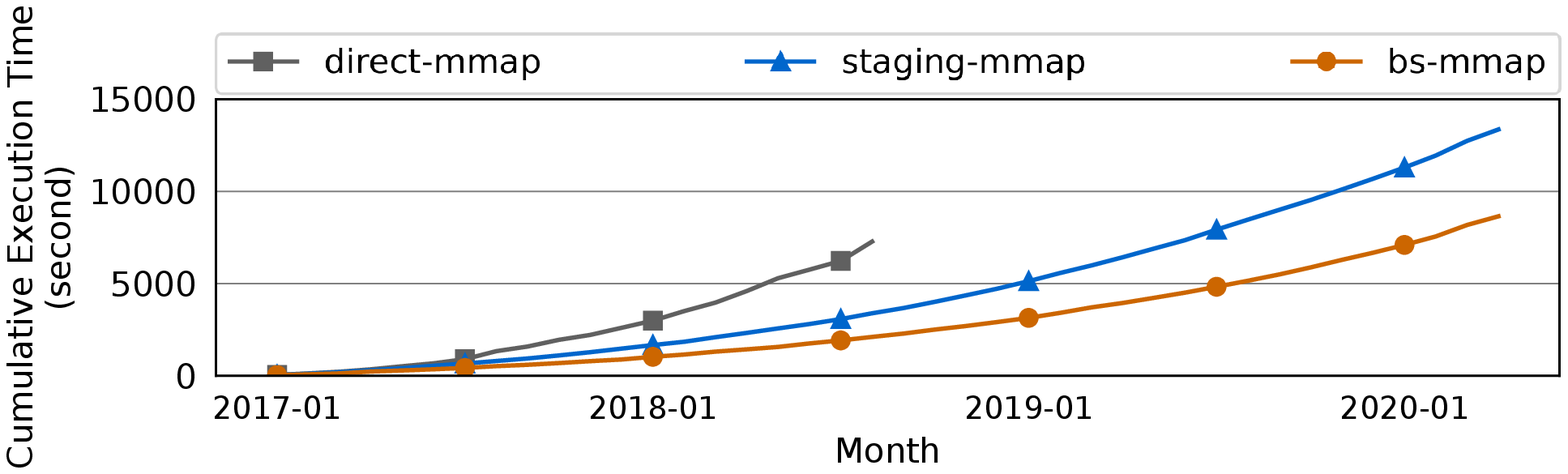}
        \caption{Reddit on VAST}
        \label{fig:reddit_vast}
    \end{subfigure}

    \caption{Cumulative execution time after each iteration (month) of constructing the Wikipedia page reference graph and the Reddit author-author graph on Lustre and VAST file systems. direct-mmap completed in only one case (Wikipedia on VAST).}
    \label{fig:incremental_graph_pagemap}
\end{figure}

\begin{figure}
    \begin{tabular}{cc}

    \begin{minipage}[t]{0.475\linewidth}
        \includegraphics[width=\textwidth]{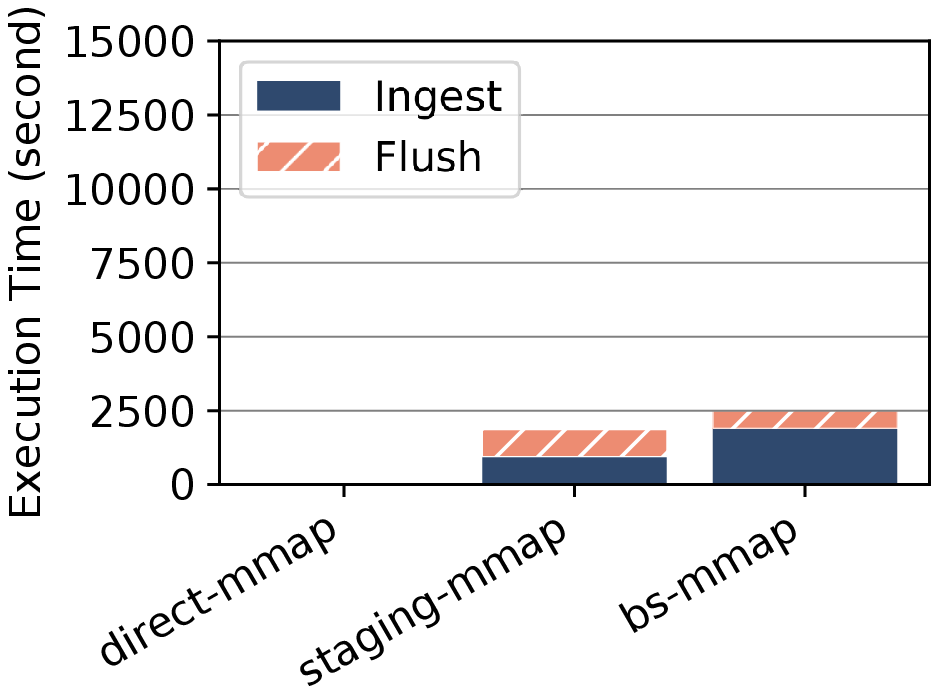}
        \subcaption{Wikipedia on Lustre}
        \label{fig:breakdown_wikipedia_lustre}
    \end{minipage}
    \begin{minipage}[t]{0.475\linewidth}
        \centering
        \includegraphics[width=\textwidth]{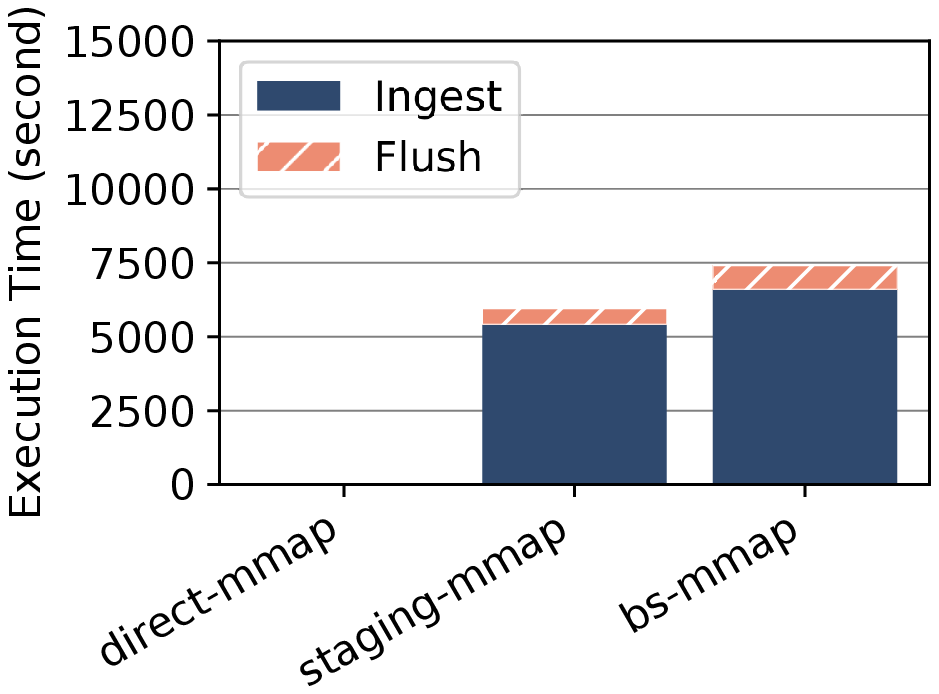}
        \subcaption{Reddit on Lustre}
        \label{fig:breakdown_reddit_lustre}
    \end{minipage}

    \\

    \begin{minipage}[c]{0.475\linewidth}
        \includegraphics[width=\textwidth]{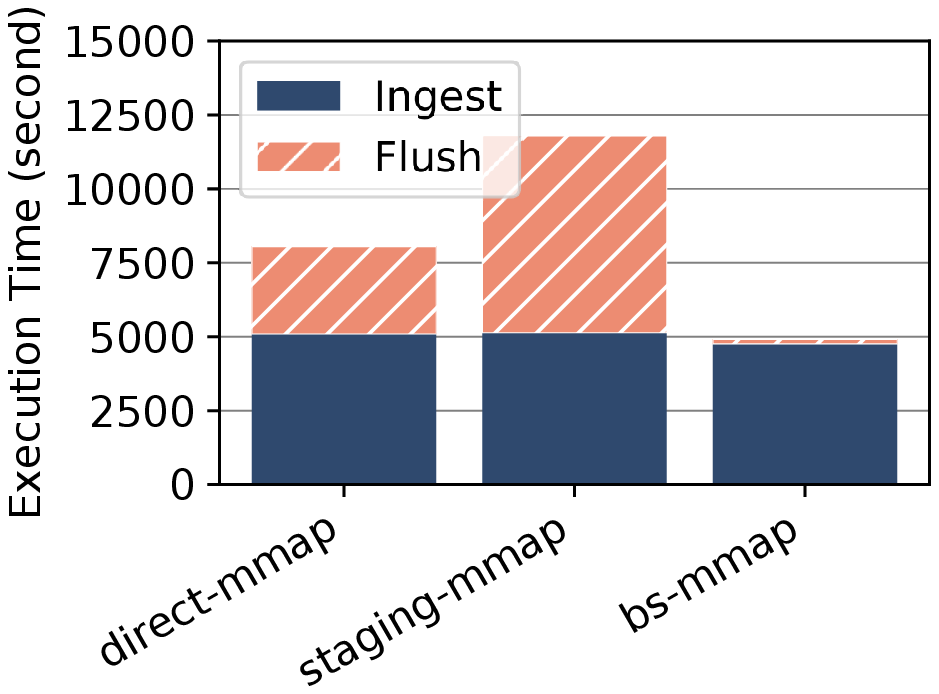}
        \subcaption{Wikipedia on VAST}
        \label{fig:breakdown_wikipedia_vast}
    \end{minipage}
    \begin{minipage}[c]{0.475\linewidth}
        \includegraphics[width=\textwidth]{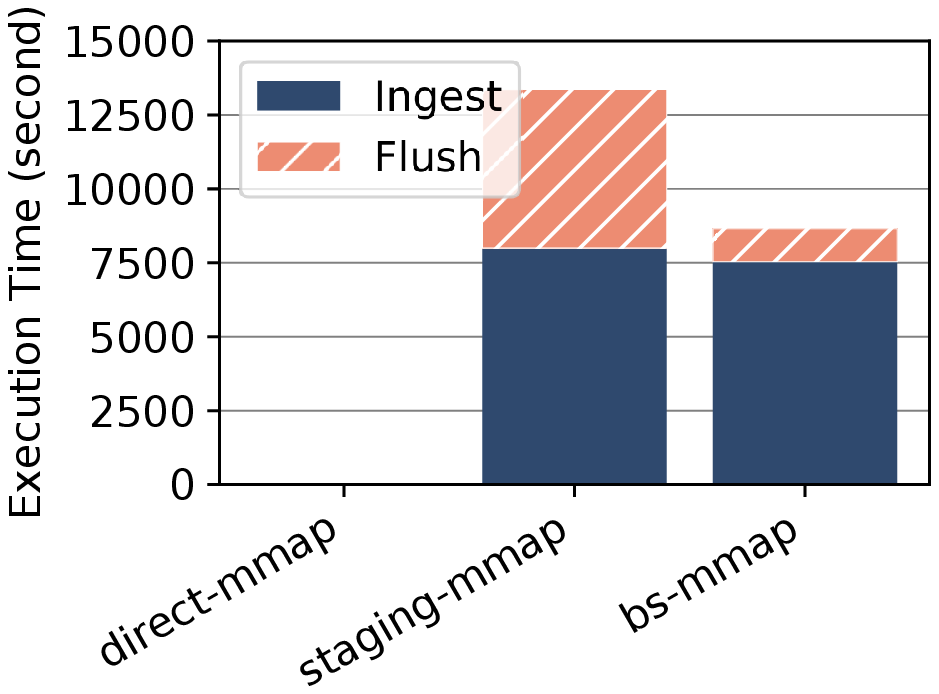}
        \subcaption{Reddit on VAST}
        \label{fig:breakdown_reddit_vast}
    \end{minipage}

    \end{tabular}

    \caption{Total time for incrementally constructing the Wikipedia page reference graph and the Reddit author-author graph on Lustre and VAST file systems. The time is broken down into total ingestion time and total flush time. direct-mmap completed in only one case (Wikipedia on VAST).}
    \label{fig:incremental_graph_breakdown}
\end{figure}

Figure~\ref{fig:incremental_graph_pagemap} contains the cumulative execution time after each iteration of constructing the Wikipedia and the Reddit graphs on Lustre and VAST.
Figure~\ref{fig:incremental_graph_breakdown} shows the time breakdown into ingestion time and flush time for each configuration.
We added staging-mmap's copy-in time to the ingestion time and copy-out time to the flush time.

First, the direct-mmap did not complete within a reasonable time except for the Wikipedia graph on VAST.
As direct-mmap needed to issue a lot of fine-grained write-backs on the fly for evacuating dirty pages to the file systems over the networks, it showed the notable slow performance, especially on Lustre.

Second, on Lustre, staging-mmap showed the best performance for both graphs. Compared with bs-mmap, it was 1.3X and 1.5X faster for the Wikipedia graph and the Reddit graph, respectively.
As the Lustre system has high bandwidth, staging-mmap was able to conduct the staging of the whole datastore to/from the local memory (tmpfs) efficiently.
We also attribute the slower ingest times of bs-mmap to the cost of accessing the file metadata of the Lustre system.
Consequently, staging-mmap was the best on the Lustre.

Third, on the VAST, bs-mmap yielded the best performance out of all three configurations. It showed 1.6X and 2.4X better performance than direct-mmap and staging-mmap for the Wikipedia graph, respectively.
bs-mmap was also 1.5X better than staging-mmap for the Reddit graph.
staging-mmap suffered from the low bandwidth of the filesystem and took considerably long times for staging out the datastores, which are included in flush time in Figure~\ref{fig:incremental_graph_breakdown}.
On the other hand, bs-mmap was able to finish the flush step quickly as it writes back only dirty pages.

\section{Application Case Study: GraphBLAS Template Library (GBTL)}
\label{sec:gbtl}

In this section, we demonstrate the high adaptability of Metall by integrating it into GraphBLAS Template Library (GBTL)~\cite{GBTL}.
In this work, we present GraphBLAS as a real application use case to demonstrate Metall persistent memory allocator benefits.
We show an example of how storing and reattaching graph containers using Metall,  eliminates the need for graph reconstruction at a one-time cost of reattaching to Metall datastore.

\subsection{Graph Analytics and GraphBLAS}

Graph analytics enables us to develop new data processing capabilities. One of the main problems in graph analytics is the need to persist the data beyond the scope of a single execution.
Graph construction, indexing, and regular updates are often more expensive than the analytics itself.
This has been observed in, for instance, large genome assembly~\cite{genome} and kNN graphs~\cite{knn}.
With persistent memory, data structures, once constructed, can be re-analyzed and updated beyond the lifetime of a single execution.
GraphBLAS specifies a set of building blocks for computing on graphs and graph-structured data, expressed in the language of linear algebra~\cite{GraphBLAS}. This approach represents graphs as sparse matrices and operations using an extended algebra of semirings. An almost unlimited variety of operators and types are supported for creating a wide range of graph algorithms. The GraphBLAS Template Library (GBTL) is a C++ reference implementation of the GraphBLAS specification \cite{GBTL}.

\subsection{Graph Analytics using GBTL}

We show a typical workflow of GBTL in Code~\ref{list:gbtl-original}.

\begin{itemize}
\item First, read edge list data stored in text files, counting the numbers of vertices and edges in the edge list;
\item Second, allocate a graph object using the information about the numbers of vertices and edges;
\item Third, ingest the input edge list into the graph object;
\item Finally, run graph algorithm(s).
\end{itemize}

As GBTL uses a normal/transient memory allocator, one has to repeat the whole graph construction step every time when running a graph algorithm.


\subsection{Integrating Metall into GBTL}

Here, we describe the work we performed for integrating Metall into GBTL.


\subsubsection{Graph Data Structure}

GBTL employs an adjacency list data structure to store the vertex and edge lists.
Its adjacency list uses vector containers internally.
We adapted the data structures to take a custom STL allocator instead of using the default one in the vector container.
We were able to complete the adaption just following the C++ standard style of implementing an allocator-aware class.
In fact, the modified data structures do not contain any code that depends on Metall.



\subsubsection{Graph Algorithm Implementation}

GBTL has a set of high-level graph algorithms built on top of GraphBLAS.
We selected five algorithms implementations to investigate the necessary change for integrating Metall into GBTL: breadth-first search, page rank, single-source-shortest-paths, triangle counting, and K-Truss.

It turned out that the only additional requirement in 'metallizing' GBTL was modifying the template parameters of the graph algorithm functions so that the functions can take a graph type with a custom allocator --- no changes were made inside the graph algorithm functions.

In addition, we found that GBTL implementations use temporary graph containers to store intermediate results in computing graph algorithms.
Specifically, we found lines like ``Graph\_t tmp\_g;'' (Graph\_t is a graph type) in multiple graph analytics functions.
Such temporary graphs need not be allocated in the persistent store and can be left as a non-persistent data structure in DRAM.

To make this convenient, we implemented another STL-compatible allocator, called \emph{fallback allocator adaptor}.
The fallback allocator adaptor \emph{fallbacks} to a normal memory allocator (e.g., malloc()) if its default constructor is called.
Metall's STL-compatible allocator (not the fallback adaptor) has to be constructed with a parameter so that it can communicate with a Metall manager object.
Thus, fallback allocator adaptor knows that the application wants to allocate the object into DRAM rather than persistent memory if no argument is passed to its constructor.
The purpose of this adaptor is to provide a way to quickly integrate Metall into an application that occasionally wants to allocate 'metallized' classes as non-persistent data structures in DRAM.
By introducing the adaptor, we were able to support Metall with only the changes in the graph data structures and a few helper functions.

\subsubsection{Graph Analytics with Metall and GBTL}

Finally, we show how the original graph analytics code (Code~\ref{list:gbtl-original}) should be changed to use Metall (Code~\ref{list:gbtl-metall}).
Specifically, we applied the following changes:
\begin{itemize}
\item Use Metall to allocate the graph at Step 2 (line 9--10).
\item Add a graph reattach mode (line 16--17).
\end{itemize}

As shown in Code~\ref{list:gbtl-metall}, Metall scopes provide a way to exit the program and reattach to the previously created data, avoiding construction time. This would be helpful to many graph analytics applications where the data structure reconstruction can be completely avoided.

\begin{lstlisting}[language=C++, caption={Workflow of a graph analytics with GBTL}, label={list:gbtl-original}]
void main() {
 // Step 1. Read edgelist
 vector<pair<int, int>> edges;
 int nv; // #of vertices
 int ne; // #of edges
 read_edge_list(``./edge_list.txt'', &edges, &nv, &ne);

 // Step 2. Allocate graph object
 auto* g = new Graph(nv, ne);

 // Step 3. Ingest edgelist to build graph
 g->build(edges);

 // Step 4. Run graph analytics
 run_analytics(*g);

 delete g;
}
\end{lstlisting}

\begin{lstlisting}[language=C++, caption={Workflow of a graph analytics with GBTL and Metall. This code is based on Code~\ref{list:gbtl-original}.}, label={list:gbtl-metall}]
using Graph_t = Graph<metall::allocator<std::byte>>;
void main() {
 Graph_t *g;
 metall::manager *mgr;
 if (create_new) { // Create a new graph
  // 1) Read edgelist, no change (code is omitted)

  // 2) Allocate graph object
  mgr = new metall::manager(metall::create_only, "/ssd/graph");
  g = mgr.construct<Graph_t>("g")(nv, ne, mgr.get_allocator());

  // 3) Ingest edgelist to build graph, no change
  g->build(edges);
 } else {
  // Re-attach graph
  mgr = new metall::manager(metall::open_read_only, "/ssd/graph");
  g = mgr.find<Graph_t>("g").first;
 }
 // 4)  Run graph analytics, no change
 run_analytics(*g);

 delete mgr;
}
\end{lstlisting}

\subsection{Demonstration}


\begin{figure}
    \centering
    \includegraphics[width=0.225\textwidth]{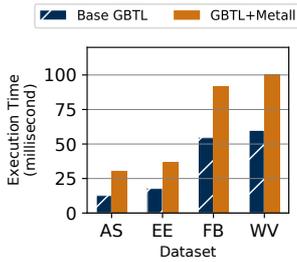}
    \caption{Graph construction time on the EPYC machine with four graph datasets. Base GBTL ran on DRAM. GBTL+Metall ran on the NVMe SSD.}
    \label{fig:gbtl-const}
\end{figure}

\begin{figure}
  \begin{minipage}[b]{0.45\linewidth}
        \centering
        \includegraphics[width=\textwidth]{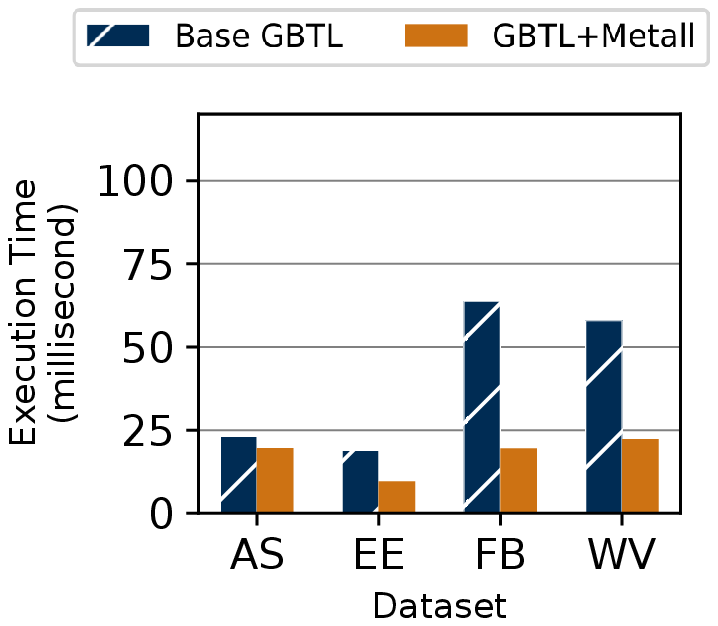}
        \subcaption{BFS}
        \label{fig:gbtl-bfs}
  \end{minipage}
  \begin{minipage}[b]{0.45\linewidth}
        \centering
        \includegraphics[width=\textwidth]{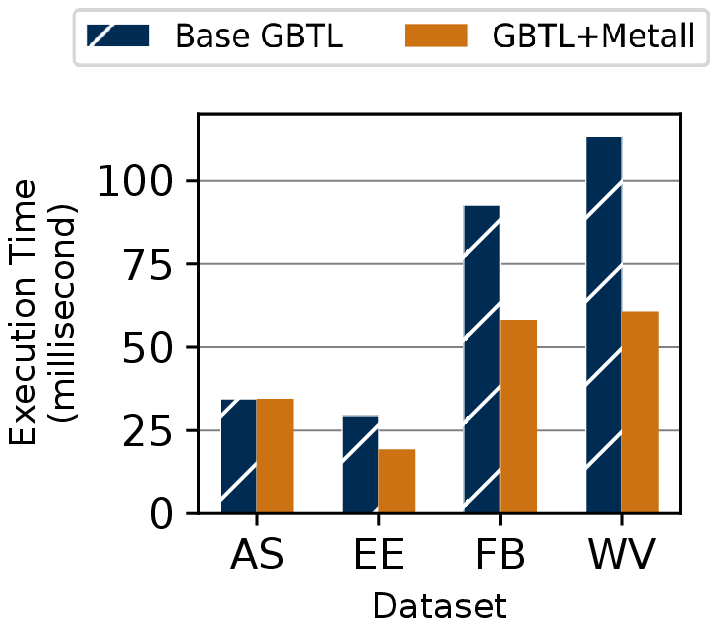}
        \subcaption{Page Rank (PR)}
        \label{fig:gbtl-pr}
  \end{minipage}
    \caption{Graph analytic time on the EPYC machine using four graph datasets. Base GBTL ran on DRAM and includes a graph construction time. GBTL+Metall ran on the NVMe SSD and includes time of reattaching a prevously constructed graph.}
    \label{fig:gbtl-algo}
\end{figure}

We use the breadth-first search (BFS) and page rank (PR) implemented in GBTL as a real application use case to demonstrate Metall persistent memory allocator benefits.
BFS is a traversal algorithm that starts at a given source vertex and produces a list of vertices that are reachable from the source vertex by traversing the edges of the graph. In each iteration, only the vertices adjacent to a newly discovered vertex are processed. Page rank is a variant of the eigenvector centrality algorithm, which measures the influence of a node in a network. It gives a rough estimate of measuring the importance of website pages.

We used four datasets from SNAP~\cite{snapnets}: as-733 (AS)~\cite{as20000102}, email-Eu-core (EE)~\cite{email-Eu-core}, ego-Facebook (FB)~\cite{snap-facebook}, and wiki-Vote (WV)~\cite{wiki-vote}.
Those graphs contain 1K--7K vertices and 14K--104K edges.

We first show the graph construction time of the original GBTL (\emph{Base GBTL}) and our 'metallized' GBTL (\emph{GBTL+Metall}) in Figure~\ref{fig:gbtl-const}.
Base GBTL constructs a graph on DRAM, whereas GBTL+Metall does that on the NVMe SSD so that it can reuse the graph data later.
GBTL+Metall was roughly 2X slower than  Base GBTL.
This is mainly due to the graph being constructed on the SSD device.

Next, we performed BFS and page rank using the two GBTL implementations (Figure~\ref{fig:gbtl-algo}).
The Base GBTL cases include graph construction time as Base GBTL needs to construct a graph from scratch every time launching a graph analytics program.
On the other hand, GBTL+Metall avoids the reconstruction time by simply reattaching a previously constructed graph.
Thus, the GBTL+Metall cases include only graph reattaching time (graphs were stored in the NVMe SSD).
GBTL+Metall achieved 3.5X better BFS execution time over Base GBTL (Figure~\ref{fig:gbtl-bfs}).
Metall's real benefit comes into the picture when multiple analytics run on the same previously constructed graph data, completely avoiding the graph reconstruction time.
Similar results were observed from the page rank analytic time in Figure~\ref{fig:gbtl-pr}.

By integrating Metall in GBTL, we were able to avoid the heavy graph construction time. This capability will be helpful to many graph analytics applications where the data structure reconstruction can be completely avoided.

Memory-mapped persistent pre-built data structures are helpful in enabling interactive real-time data science applications with large persistent data structures in unprecedented scales of data without going through traditional serialization and data structure reconstruction. Application developers can create custom complex persistent and consistent data structures. This ability to attach and detach from previously created datasets in a lightweight manner gives a powerful workflow software productivity benefit.

\FloatBarrier

\section{Related Work}
\label{sec:related_work}

\subsection{Large-scale Graph Processing with Persistent Memory}
Many studies have been conducted for large-scale graph processing on persistent memory using mmap and showed notable performance~\cite{Pearce2010, Iwabuchi2016}, including evaluating large-scale graph processing on Intel Optane DC Persistent Memory (e.g.,~\cite{Gurbinder,peng2019}).
Metall is aiming at helping application developers with respect to memory allocation and data management.

\subsection{Memory Allocator}
Boost.Interprocess (BIP)~\cite{Boost} offers higher-level allocator mechanisms on top of a file-backed mmap mechanism.
However, as it is not designed intentionally as a persistent allocator, there are some drawbacks.
For example,
1) BIP uses a single tree to manage memory allocations --- such design will suffer from many allocations and not scale well with multiple threads due to lock contention;
2) it is not capable of deallocating file (persistent memory) space.
On the other hand, as Metall is not designed for interprocess communication, it does not support process synchronization.
Except for the restriction, applications that already work with the allocators in Boost.Interprocess (especially managed\_mapped\_file allocates) should work with Metall without modification.

NVMalloc~\cite{NVMalloc} enables applications to allocate memory on a distributed non-volatile memory (NVM) storage system.
NVMalloc creates a file per memory allocation request, but it does not have a mechanism of aggregating multiple allocations into a single file
--- creating a file and mapping it to the main memory are expensive operations; therefore, NVMalloc will not be suitable when the application causes many relatively small allocations.

Persistent Memory Development Kit (PMDK)~\cite{pmemio} is a collection of libraries focusing on byte-addressable persistent memory.
libpmemobj in PMDK provides a persistent memory allocator like Metall with a fine-grained persistence policy (see Section~\ref{sec:persistence_policy}).
nvm\_malloc~\cite{nvm_malloc} is another work for byte-addressable persistent memory with a fine-grained persistence policy.
Ralloc~\cite{Ralloc} also targets byte-addressable persistent memory.
It supports failure-atomic memory allocations by asking applications to write a function that traverses all active pointers.
Ralloc possesses better performance over libpmemobj~\cite{Ralloc}.
Compared with those works, Metall is designed for both block devices and byte-addressable persistent memory so that applications can utilize a wide range of persistent memory technologies in their environment.
We also demonstrated that Metall showed competitive performance with Ralloc (Section~\ref{sec:evaluation}).

Several studies have been conducted and shown remarkable performance as for heap memory allocator,
e.g., jemalloc~\cite{jemalloc} and Supermalloc~\cite{Supermalloc}.
However, unfortunately, it is not trivial to extend those allocators for persistent memory because allocators themselves also have to be stored in persistent memory so that they can resume the previous status.

\subsection{System Software}
Several projects investigated mmap technology.
For example, DI-MMAP~\cite{DI-MMAP} improved mmap page cache performance; UMap is a user-level mmap library that lets users control the page cache policy more flexibly~\cite{UMap}.

Failure-atomic msync() (FAMS)~\cite{FAMS} is a mechanism that guarantees that the backing file of a mmap() region always reflects the most recent successful msync(), regardless of crashes.
FAMS could be useful to enhance Metall's failure atomic support.
There are several FAMS implementations.
For example, it is implemented in NOVA filesystem~\cite{NOVA}.
famus is a user-level FAMS library~\cite{famus}.

\subsection{Persistent Data Store}
In terms of storing data persistently, a key-value store is a popular database model and is designed to scale to a very large size easily.
(for example, LevelDB~\cite{LevelDB}, RocksDB~\cite{RocksDB}, and MongoDB~\cite{MongoDB} use mmap).
Metall's key benefit is that it allows for custom data structures to be stored, not just key-value pairs.

Hierarchical Data Format (HDF)~\cite{HDF}, particularly HDF5, has been used in many large-scale data analytics applications.
It allows applications to store data in portable formats.
On the other hand, Metall is designed as a lightweight data store library by limiting data portability.

\section{Conclusion}
\label{sec:conclusion}

Thanks to the recent notable performance improvements, various types of NVRAM devices are already used in many HPC systems today.
We anticipate that persistent data-centric analytics will be a powerful model for accelerating next-generation large-scale data analytics.
To leverage various persistent memory devices in the current and future exascale HPC systems, we developed a persistent memory allocator Metall.
Metall allows applications to allocate data structures into persistent memory transparently with a reasonable code migration cost.

Metall achieved up to 11.7x and 48.3x performance improvements over Boost.Interprocess and memkind (PMEM kind), respectively, on the dynamic graph construction workload with the node-local conventional NVMe SSD and the node-local emerging byte-addressable persistent memory.
We demonstrated Metall's high adaptability and its benefit on the real graph processing workload using GBTL.
We also investigated and developed the techniques for improving sparse data update performance on network-attached file systems.

This study's outcomes indicate that Metall will be a strong tool for accelerating future large-scale data analytics by enabling applications to efficiently and fully leverage persistent memory.

\section*{Acknowledgment}
This work was performed under the auspices of the U.S. Department of Energy by Lawrence Livermore National Laboratory under Contract DE-AC52-07NA27344 (LLNL-JRNL-825588).
Experiments were performed at the Livermore Computing facility.

This research was supported by the Exascale Computing Project (17-SC-20-SC),
a collaborative effort of the U.S. Department of Energy Office of Science and the National Nuclear Security Administration.

\bibliographystyle{elsarticle-num}
\bibliography{metall}

\begin{thebibliography}{10}
\expandafter\ifx\csname url\endcsname\relax
  \def\url#1{\texttt{#1}}\fi
\expandafter\ifx\csname urlprefix\endcsname\relax\def\urlprefix{URL }\fi
\expandafter\ifx\csname href\endcsname\relax
  \def\href#1#2{#2} \def\path#1{#1}\fi

\bibitem{ECP}
Exascale computing project, \url{https://www.exascaleproject.org/}, (Accessed
  on 03/29/2021).

\bibitem{Reed2015}
D.~A. Reed, J.~Dongarra, \href{https://doi.org/10.1145/2699414}{Exascale
  computing and big data}, Commun. ACM 58~(7) (2015) 56--68.
\newblock \href {https://doi.org/10.1145/2699414} {\path{doi:10.1145/2699414}}.
\newline\urlprefix\url{https://doi.org/10.1145/2699414}

\bibitem{Boost}
{Boost C++ Libraries}, \url{https://www.boost.org/}, (Accessed on 03/29/2021).

\bibitem{Supermalloc}
B.~C. Kuszmaul, {SuperMalloc}: A super fast multithreaded malloc for 64-bit
  machines, in: Proceedings of the 2015 International Symposium on Memory
  Management, ISMM '15, ACM, 2015, pp. 41--55.
\newblock \href {https://doi.org/10.1145/2754169.2754178}
  {\path{doi:10.1145/2754169.2754178}}.

\bibitem{GBTL}
{GraphBLAS Template Library (GBTL), v. 3.0},
  \url{https://github.com/cmu-sei/gbtl}, (Accessed on 03/29/2021).

\bibitem{memkind}
Memkind, \url{http://memkind.github.io/memkind/}, (Accessed on 03/29/2021).

\bibitem{JSFI162}
J.~L{\"u}ttgau, M.~Kuhn, K.~Duwe, Y.~Alforov, E.~Betke, J.~Kunkel, T.~Ludwig,
  \href{https://www.superfri.org/superfri/article/view/162}{Survey of storage
  systems for high-performance computing}, Supercomputing Frontiers and
  Innovations 5~(1) (2018).
\newline\urlprefix\url{https://www.superfri.org/superfri/article/view/162}

\bibitem{Hirofuchi}
T.~Hirofuchi, R.~Takano, A prompt report on the performance of intel optane dc
  persistent memory module, IEICE TRANSACTIONS on Information and Systems
  103~(5) (2020) 1168--1172.

\bibitem{Lee19}
G.~Lee, S.~Shin, W.~Song, T.~J. Ham, J.~W. Lee, J.~Jeong,
  \href{https://www.usenix.org/conference/atc19/presentation/lee-gyusun}{Asynchronous
  i/o stack: A low-latency kernel i/o stack for ultra-low latency ssds}, in:
  2019 {USENIX} Annual Technical Conference ({USENIX} {ATC} 19), {USENIX}
  Association, Renton, WA, 2019, pp. 603--616.
\newline\urlprefix\url{https://www.usenix.org/conference/atc19/presentation/lee-gyusun}

\bibitem{OptaneSSD}
{Intel Optane SSD DC}, \url{https://www.intel.com/} (August 2021).

\bibitem{Skyway}
K.~Nguyen, L.~Fang, C.~Navasca, G.~Xu, B.~Demsky, S.~Lu,
  \href{https://doi.org/10.1145/3173162.3173200}{Skyway: Connecting managed
  heaps in distributed big data systems}, in: Proceedings of the Twenty-Third
  International Conference on Architectural Support for Programming Languages
  and Operating Systems, ASPLOS '18, Association for Computing Machinery, New
  York, NY, USA, 2018, pp. 56--69.
\newblock \href {https://doi.org/10.1145/3173162.3173200}
  {\path{doi:10.1145/3173162.3173200}}.
\newline\urlprefix\url{https://doi.org/10.1145/3173162.3173200}

\bibitem{pmemio}
{pmem.io Persistent Memory Programming}, \url{https://pmem.io/}, (Accessed on
  03/29/2021).

\bibitem{MOD}
S.~Haria, M.~D. Hill, M.~M. Swift,
  \href{https://doi.org/10.1145/3373376.3378472}{{MOD}: Minimally ordered
  durable datastructures for persistent memory}, in: Proceedings of the
  Twenty-Fifth International Conference on Architectural Support for
  Programming Languages and Operating Systems, ASPLOS '20, Association for
  Computing Machinery, New York, NY, USA, 2020, pp. 775--788.
\newblock \href {https://doi.org/10.1145/3373376.3378472}
  {\path{doi:10.1145/3373376.3378472}}.
\newline\urlprefix\url{https://doi.org/10.1145/3373376.3378472}

\bibitem{reflink}
ioctl\_ficlonerange(2) - {Linux manual page},
  \url{https://man7.org/linux/man-pages/man2/ioctl_ficlonerange.2.html},
  (Accessed on 03/29/2021).

\bibitem{Ralloc}
W.~Cai, H.~Wen, H.~A. Beadle, C.~Kjellqvist, M.~Hedayati, M.~L. Scott,
  \href{https://doi.org/10.1145/3381898.3397212}{Understanding and optimizing
  persistent memory allocation}, in: Proceedings of the 2020 ACM SIGPLAN
  International Symposium on Memory Management, ISMM 2020, Association for
  Computing Machinery, New York, NY, USA, 2020, pp. 60--73.
\newblock \href {https://doi.org/10.1145/3381898.3397212}
  {\path{doi:10.1145/3381898.3397212}}.
\newline\urlprefix\url{https://doi.org/10.1145/3381898.3397212}

\bibitem{jemalloc}
jemalloc, \url{http://jemalloc.net/}, (Accessed on 03/29/2021).

\bibitem{Lustre}
Lustre, \url{https://www.lustre.org/}, (Accessed on 03/29/2021).

\bibitem{uselton2013file}
A.~Uselton, N.~Wright, A file system utilization metric for i/o
  characterization, in: Proc. of the Cray User Group conference, 2013.

\bibitem{LinuxKernelPagemap}
Pagemap, from the userspace perspective,
  \url{https://www.kernel.org/doc/Documentation/vm/pagemap.txt}, (Accessed on
  03/29/2021).

\bibitem{VAST}
{Exascale NAS} --- whitepaper || {VAST Data},
  \url{https://vastdata.com/exascale-nas-whitepaper/}, (Accessed on
  03/29/2021).

\bibitem{R-MAT}
D.~Chakrabarti, Y.~Zhan, C.~Faloutsos, {R-MAT}: A recursive model for graph
  mining, in: Proceedings of the 2004 SIAM International Conference on Data
  Mining, SIAM, 2004, pp. 442--446.

\bibitem{genome}
C.~{Boucher}, A.~{Bowe}, T.~{Gagie}, S.~{Puglisi}, K.~{Sadakane},
  Variable-order de bruijn graphs, in: 2015 Data Compression Conference, 2015.
\newblock \href {https://doi.org/10.1109/DCC.2015.70}
  {\path{doi:10.1109/DCC.2015.70}}.

\bibitem{knn}
J.~{Chen}, H.~{Fang}, Y.~{Saad}, Fast approximate knn graph construction for
  high dimensional data via recursive lanczos bisection, in: Journal of Machine
  Learning Research 10 (2009), 2009.

\bibitem{GraphBLAS}
J.~{Kepner}, P.~{Aaltonen}, D.~{Bader}, A.~{Bulu{\c c}}, F.~{Franchetti},
  J.~{Gilbert}, D.~{Hutchison}, M.~{Kumar}, A.~{Lumsdaine}, H.~{Meyerhenke},
  S.~{McMillan}, C.~{Yang}, J.~D. {Owens}, M.~{Zalewski}, T.~{Mattson},
  J.~{Moreira}, Mathematical foundations of the graphblas, in: 2016 IEEE High
  Performance Extreme Computing Conference (HPEC), 2016, pp. 1--9.
\newblock \href {https://doi.org/10.1109/HPEC.2016.7761646}
  {\path{doi:10.1109/HPEC.2016.7761646}}.

\bibitem{snapnets}
J.~Leskovec, A.~Krevl, {SNAP Datasets}: {Stanford} large network dataset
  collection, \url{http://snap.stanford.edu/data} (Jun. 2014).

\bibitem{as20000102}
J.~Leskovec, J.~Kleinberg, C.~Faloutsos, Graphs over time: densification laws,
  shrinking diameters and possible explanations, in: Proceedings of the
  eleventh ACM SIGKDD international conference on Knowledge discovery in data
  mining, 2005, pp. 177--187.

\bibitem{email-Eu-core}
J.~Leskovec, J.~Kleinberg, C.~Faloutsos, Graph evolution: Densification and
  shrinking diameters, in: ACM Transactions on Knowledge Discovery from Data
  (ACM TKDD), 2007.

\bibitem{snap-facebook}
M.~J, J.~Leskovec, Learning to discover social circles in ego networks, in:
  Proceedings of the The Neural Information Processing Systems 2012, 2012.

\bibitem{wiki-vote}
J.~Leskovec, D.~Huttenlocher, J.~Kleinberg, Signed networks in social media,
  in: Proceedings of the international conference of Human-Computer Interaction
  2010, 2010.

\bibitem{Pearce2010}
R.~Pearce, M.~Gokhale, N.~M. Amato, Multithreaded asynchronous graph traversal
  for in-memory and semi-external memory, in: Proceedings of the 2010 ACM/IEEE
  International Conference for High Performance Computing, Networking, Storage
  and Analysis, SC '10, IEEE Computer Society, 2010, pp. 1--11.
\newblock \href {https://doi.org/10.1109/SC.2010.34}
  {\path{doi:10.1109/SC.2010.34}}.

\bibitem{Iwabuchi2016}
K.~{Iwabuchi}, S.~{Sallinen}, R.~{Pearce}, B.~V. {Essen}, M.~{Gokhale},
  S.~{Matsuoka}, Towards a distributed large-scale dynamic graph data store,
  in: 2016 IEEE International Parallel and Distributed Processing Symposium
  Workshops (IPDPSW), 2016, pp. 892--901.
\newblock \href {https://doi.org/10.1109/IPDPSW.2016.189}
  {\path{doi:10.1109/IPDPSW.2016.189}}.

\bibitem{Gurbinder}
G.~Gill, R.~Dathathri, L.~Hoang, R.~Peri, K.~Pingali,
  \href{http://arxiv.org/abs/1904.07162}{Single machine graph analytics on
  massive datasets using {Intel Optane DC Persistent Memory}}, CoRR
  abs/1904.07162 (2019).
\newblock \href {http://arxiv.org/abs/1904.07162} {\path{arXiv:1904.07162}}.
\newline\urlprefix\url{http://arxiv.org/abs/1904.07162}

\bibitem{peng2019}
I.~B. Peng, M.~B. Gokhale, E.~W. Green, System evaluation of the {Intel Optane}
  byte-addressable {NVM}, arXiv preprint arXiv:1908.06503 (2019).

\bibitem{NVMalloc}
C.~{Wang}, S.~S. {Vazhkudai}, X.~{Ma}, F.~{Meng}, Y.~{Kim}, C.~{Engelmann},
  {NVMalloc}: Exposing an aggregate {SSD} store as a memory partition in
  extreme-scale machines, in: 2012 IEEE 26th International Parallel and
  Distributed Processing Symposium, 2012, pp. 957--968.
\newblock \href {https://doi.org/10.1109/IPDPS.2012.90}
  {\path{doi:10.1109/IPDPS.2012.90}}.

\bibitem{nvm_malloc}
D.~Schwalb, T.~Berning, M.~Faust, M.~Dreseler, H.~Plattner, nvm{\_}malloc:
  Memory allocation for {NVRAM}., ADMS@ VLDB 15 (2015) 61--72.

\bibitem{DI-MMAP}
B.~V. {Essen}, H.~{Hsieh}, S.~{Ames}, M.~{Gokhale}, {DI-MMAP}: A high
  performance memory-map runtime for data-intensive applications, in: 2012 SC
  Companion: High Performance Computing, Networking Storage and Analysis, 2012,
  pp. 731--735.
\newblock \href {https://doi.org/10.1109/SC.Companion.2012.99}
  {\path{doi:10.1109/SC.Companion.2012.99}}.

\bibitem{UMap}
I.~{Peng}, M.~{McFadden}, E.~{Green}, K.~{Iwabuchi}, K.~{Wu}, D.~{Li},
  R.~{Pearce}, M.~{Gokhale}, Umap: Enabling application-driven optimizations
  for page management, in: 2019 IEEE/ACM Workshop on Memory Centric High
  Performance Computing (MCHPC), 2019, pp. 71--78.
\newblock \href {https://doi.org/10.1109/MCHPC49590.2019.00017}
  {\path{doi:10.1109/MCHPC49590.2019.00017}}.

\bibitem{FAMS}
T.~Kelly, \href{http://doi.acm.org/10.1145/3358955.3358957}{Persistent memory
  programming on conventional hardware}, Queue 17~(4) (2019) 10:1--10:20.
\newblock \href {https://doi.org/10.1145/3358955.3358957}
  {\path{doi:10.1145/3358955.3358957}}.
\newline\urlprefix\url{http://doi.acm.org/10.1145/3358955.3358957}

\bibitem{NOVA}
J.~Xu, S.~Swanson,
  \href{https://www.usenix.org/conference/fast16/technical-sessions/presentation/xu}{{NOVA}:
  A log-structured file system for hybrid volatile/non-volatile main memories},
  in: 14th {USENIX} Conference on File and Storage Technologies ({FAST} 16),
  {USENIX} Association, Santa Clara, CA, 2016, pp. 323--338.
\newline\urlprefix\url{https://www.usenix.org/conference/fast16/technical-sessions/presentation/xu}

\bibitem{famus}
famus: Failure-atomic msync() in user space,
  http://web.eecs.umich.edu/~tpkelly/famus/.

\bibitem{LevelDB}
Level{DB}, \url{https://github.com/google/leveldb}, (Accessed on 03/29/2021).

\bibitem{RocksDB}
{RocksDB: A Persistent Key-Value Store for Flash and RAM Storage},
  \url{https://github.com/facebook/rocksdb}, (Accessed on 03/29/2021).

\bibitem{MongoDB}
K.~Chodorow, {MongoDB}: the definitive guide: powerful and scalable data
  storage, " O'Reilly Media, Inc.", 2013.

\bibitem{HDF}
{The HDF Group - ensuring long-term access and usability of HDF data and
  supporting users of HDF technologies}, \url{https://www.hdfgroup.org/},
  (Accessed on 02/20/2021).

\end{thebibliography}


\end{document}